\documentclass[pra,twocolumn,showpacs,nobibnotes,floatfix,superscriptaddress]{revtex4}

\usepackage{latexsym,amsmath,amssymb,amsfonts,mathbbol,graphicx,color}
\usepackage{dcolumn}
\usepackage{bm}

\begin{document}

\title{Entanglement Dynamics in Three Qubit $X$-States}

\author{Yaakov S. Weinstein}
\affiliation{Quantum Information Science Group, {\sc Mitre},
260 Industrial Way West, Eatontown, NJ 07224, USA}


\begin{abstract}
I explore the entanglement dynamics of a three qubit system in an initial $X$-state 
undergoing decoherence including the possible exhibition of entanglement sudden death (ESD). 
To quantify entanglement I utilize negativity measures and make use of appropriate 
entanglement witnesses. The negativity results are then extended to $X$-states 
with an arbitraty number of qubits. I also demonstrate non-standard behavior of the 
tri-partite negativity entanglement metric, its sudden appearance after some amount 
of decoherence followed quickly by its disappearance.  Finally, I solve for a lower bound on 
the three qubit $X$-state concurrence, demonstrate when this bound goes to zero, 
and outline simplifcations for the calculation of higher order $X$-state concurrences.
\end{abstract}


\maketitle

\section{Introduction}

Entanglement is a quantum mechanical phenomenon in which quantum systems exhibit 
correlations above and beyond what is classically possible. As such it is a crucial
resource for many aspects of quantum information processing 
including quantum computation, quantum cryptography and communications, and 
quantum metrology \cite{book}. Due to its fundamental, and increasingly practical, 
importance there is a growing body of literature dedicated to studies of entanglement. 
Nevertheless, many aspects of entanglement, especially 
multi-partite entanglement and its evolution, are in need of further 
exploration \cite{HHH}.

The unavoidable degradation of entanglement due to decoherence has severely
hampered experimental attempts to realize quantum information protocols. 
Decoherence is a result of unwanted interactions between the system of interest
and its environment. Highly entangled, and thus highly non-classical, states 
may be severely corrupted by decoherence  \cite{Dur}. This is especially 
troubling as these states tend to be the most potentially useful for quantum information
protocols. An extreme manifestation of the detrimental effects of decoherence on 
entanglement is entanglement suddent death (ESD): in which decoherence causes 
a complete loss of entanglement in a finite time \cite{DH,YE1} despite the fact that 
the system coherence goes to zero only asymptotically. Much has been written about this 
aspect of entanglement for bi-partite systems and there have been several initial 
experimental studies of this phenomenon \cite{expt}. Fewer studies look at 
ESD, and specifically ESD with respect to multi-partite entanglement, in multi-partite systems 
\cite{ACCAD,LRLSR,YYE,BYW,YSW2}. 

A class of two qubit states that are generally known to exhibit ESD are the so called 
$X$-states \cite{xstate}, so named due to the pattern  of non-zero density 
matrix elements. These states play an important role in a number of physical 
systems \cite{Xexpt}, and allow for easy calculation of certain entanglement measures. 
In this paper, I explore the entanglement dynamics of three qubit 
$X$-states in dephasing and depolarizing environments as a function of decoherence strength. 
Previous studies of three qubit $X$-shaped states utilize more restrictive sets of states: 
GHZ-diagonal states \cite{GS} and generalized GHZ-diagonal states \cite{A}. Other papers have 
examined specific examples of three qubit $X$-state entanglement including the effects of 
dephasing on a three-qubit quantum error correction protocol \cite{YSW1}. 

To quantify entanglement within the three qubit systems I utilize the negativity, 
$N_j$, defined as the most negative eigenvalue of the partial 
transpose of the density matrix \cite{neg} with respect to qubit $j$. This 
provides three distinct entanglement measures. As a pure tri-partite
entanglement metric for mixed states I will use the tri-partite negativity, $N^{(3)}$
which is simply the third root of the product of the negativities with respect to each of 
three qubits \cite{SGA}, $N^{(3)} \equiv (N_1N_2N_3)^{1/3}$. A mixed state with non-zero 
$N^{(3)}$ is distillable to a GHZ state. It is important to note the existence of bound 
entanglement which may be present even if all negativity measures in a three qubit system
are equal to zero. Thus, when I refer to ESD of a state with respect to given negativity metrics 
this should not be confused with separability of the state. Nevertheless, besides general interest
in the behavior of these entanglement metrics, the disappearance of negativity plays an important 
role in quantum information protocols in that it indicates that the entanglement of the state is 
not distillable.  
 
The physical significance of $X$-states mentioned above demands and efficient means of 
experimentally determining the presence of entanglement. This can be accomplished via `entanglement witnesses.'
Three qubit states can be separated into four broad categories: separable (in all three qubits), 
biseparable, and there exist two types of locally inequivalent tri-partite entanglement (GHZ 
and W-type) \cite{DVC}. Reference \cite{ABLS} provides a similar classification schemes for 
mixed states each of which includes within it the previous classes. These are separable (S) 
states, bi-separable (B) states, W states, and GHZ states, which encompasses the complete 
set of three qubit states. 

Entanglement witnesses are used to determine in which class a given state belongs.
These observables give a positive or zero expectation value for 
all states of a given class and negative expectation values for at least one 
state in a higher ({\it i.e.}~more inclusive) class. I will make use of specific
entanglement witnesses \cite{ABLS} that will identify whether a state is in the 
GHZ$\backslash$W class ({\it i.e.}~a state in the GHZ class but not in the W class), 
in which case the state has experimentally observable GHZ-type tri-partite entanglement. 
Though the use of entanglement witness cannot guarantee that entanglement is not present, it does
give experimental bounds on whether the entanglement can be observed.

The results presented in this paper are (i) the analytical determination of various negativity measures 
for $X$-states of an arbitrary number of qubits including how the negativity evolves under decoherence, 
(ii) the demonstration that negativity disappears in finite time for $X$-states subject to different 
types of decoherence and the (analytical and numerical) determination 
of the decoherence strength when this occurs, (iii) the analytical calculation of the expectation value of 
$X$-states undergoing decoherence with respect to appropriate entanglement witnesses, (iv) the demonstration 
of the sudden appearance, only at non-zero decoherence strength, in some $X$-states of the tri-partite
negativity, and (v) the calculation of a bound on the three qubit concurrence for $X$-states and the 
description of how this can be extended to more qubits. In addition, I prove in 
the Appendix that the set of generalized GHZ-diagonal states do not cover all possible $X$-states.

\section{Three Qubit $X$-States}

There are a number of classes of three qubit states whose entanglement properties
have been studied and whose non-zero density matrix elements form an $X$ shape:
\begin{equation}
\rho_X(a_j,b_j,c_j) = \left(
\begin{array}{cccccccc}
a_1 & 0 & 0 & 0 & 0 & 0 & 0 & c_1\\
0 & a_2 & 0 & 0 & 0 & 0 & c_2 & 0\\
0 & 0 & a_3 & 0 & 0 & c_3 & 0 & 0\\
0 & 0 & 0 & a_4 & c_4 & 0 & 0 & 0\\
0 & 0 & 0 & c_4^* & b_4 & 0 & 0 & 0\\
0 & 0 & c_3^* & 0 & 0 & b_3 & 0 & 0\\
0 & c_2^* & 0 & 0 & 0 & 0 & b_2 & 0\\
c_1^* & 0 & 0 & 0 & 0 & 0 & 0 & b_1\\
\end{array}
\right)
\label{xstate}
\end{equation}
where $j = 1,...,4$. The most basic is a pure three qubit GHZ state with wavefunction 
$|\psi_k^{\pm}\rangle = \frac{1}{\sqrt{2}}(|k\rangle\pm|\overline{k}\rangle)$, where
$k$ is a three bit binary number between zero and seven and $\overline{k}$ is the result of 
flipping each bit of $k$. The density matrix of this state is $\rho_X(1/2_j,1/2_j,\pm1/2_j)$. 
Mixed states that are diagonal in the basis of these eight states form the 
set of GHZ-diagonal states studied in \cite{GS}. The basis states have coefficients 
$\sqrt{\lambda_k^{\pm}}$ for all $0 < k < 3$, the squares of which sum to one. 
The density matrix elements of these states are thus 
$a_j = b_j = \lambda_k^{+}+\lambda_k^{-}$ and $c_j = \lambda_k^{+}-\lambda_k^{-}$. 

A generalized GHZ state is a non-maximally entangled state of the form:
\begin{equation}
|\psi_k^{\pm}(\alpha,\beta)\rangle = \alpha|k\rangle\pm\beta|\overline{k}\rangle.
\end{equation}
The density matrix of this state is $\rho_X(|\alpha|^2_j,|\beta|^2_j,\pm\alpha\beta^*_j)$ 
for $j = 1,...,4$. An incoherent mixture of generalized GHZ states where $k$ now ranges from 0 to 7 
(as opposed to 0 to 3 used in in \cite{GS}) form a generalized GHZ-diagonal state. 
These states are studied in \cite{A} and have the form:
\begin{equation}
\rho = \sum_{k = 0}^7\lambda_k^+|\psi_k^+(\alpha,\beta)\rangle\langle\psi_k^+(\alpha,\beta)|+\lambda_k^-|\psi_k^-(\alpha,\beta)\rangle\langle\psi_k^-(\alpha,\beta)|.
\end{equation}
For these states the density matrix elements are as follows:
\begin{eqnarray}
a_j &=& |\alpha|^2(\lambda_k^{+}+\lambda_k^{-}) + |\beta|^2(\lambda_{\overline{k}}^{+}+\lambda_{\overline{k}}^{-}) \\
b_j &=& |\beta|^2(\lambda_k^{+}+\lambda_k^{-}) + |\alpha|^2(\lambda_{\overline{k}}^{+}+\lambda_{\overline{k}}^{-}) \\
c_j &=& \alpha\beta^*(\lambda_k^{+}-\lambda_k^{-}) + \alpha^*\beta(\lambda_{\overline{k}}^{+}-\lambda_{\overline{k}}^{-}) 
\end{eqnarray}
for $1 < j < N/2$ and $k = j - 1$. However, as shown in the Appendix,
generalized GHZ-diagonal states do not include all possible $X$-states. This is due to the restriction 
of constant $\alpha$ and $\beta$ for all contributing generalized GHZ states. 

In this paper I consider $X$-states that are completely general, limited only by the restriction 
that the state is a proper density matrix. For convenience, I will refer to the four density matrix elements 
$a_j, b_j, c_j$ and $c_j^*$ of the X-state as a GHZ-type state. At most, four GHZ-type states contribute to 
each three qubit $X$-state.

\section{X-State Entanglement}

An $X$-state is a mixed state that can be written as a sum of GHZ-type states. 
When a partial trace is taken over any one of the three qubits of an $X$-state the resulting 
two qubit matrix is diagonal. This demonstrates that the entanglement of an $X$-state is either
tri-partite or biseparable but not completely separable. Before calculating any specific
entanglement metric and studying its decay in a given decohering environment, we note that 
an upper bound on the entanglement decay was derived in \cite{A} for a number of different 
decohering environments. Though these bounds were calculated for the more limited generalized 
GHZ-diagonal states they appear to be appropriate to the states studied in this work. However, 
these upper bounds go to zero only in the limit of complete decoherence. Thus, the states never 
exhibit entanglement sudden death for any entanglement metric and the bounds cannot be used to 
study the ESD phenomenon. Below, I explore specific entanglement metrics for which I provide 
analytical solutions to exactly calculate the decoherence strength at which the $X$-states 
exhibit ESD for the given entanglement metrics. While ESD of these metrics cannot guarantee
separability of the $X$-state it does provide important information concerning distillability 
and the ability to determine the presence of entanglement. 

To calculate the negativity of a three qubit $X$-state we take the eigenvalues 
of the partial transpose of the density matrix with respect to one of the qubits. 
These 24 eigenvalues (8 for each possible partial transpose) are all of the form:
\begin{equation}
E_{ij} = \frac{1}{2}\left(a_j+b_j\pm\sqrt{(a_j-b_j)^2+4|c_i|^2}\right)
\label{XEigs}
\end{equation}
for all $i,j = 1,...,4$ and $i\neq j$. From these eigenvalues 
one can see how the negativity detects the entanglement of an $X$-state. 
Let us first assume a GHZ-type state with the only non-zero elements 
$a_j, b_j, c_j$ and $c_j^*$. The eigenvalues which utilize elements 
$a_j$ and $b_j$ cannot be negative (since $a_j+b_j = 1$ and $c_i = 0$ for 
all $i\neq j$). An additional three eigenvalues will be equal to $-|c_j|$, 
demonstrating the entanglement in the system. $X$-states 
that are sums of two GHZ type states have non-zero elements 
$a_i, a_j, b_i, b_j, c_i, c_j, c_i^*, c_j^*$. Such states will again 
have negative eigenvalues $-|c_i|$, $-|c_j|$ and two additional possibly negative eigenvalues 
$\frac{1}{2}(a_k+b_k-\sqrt{(a_k-b_k)^2+4|c_\ell|^2})$ where $k,\ell = i,j$ and 
$k \neq \ell$. As more density matrix elements of the $X$-state are filled up  
the eigenvalues tend to have the form of these latter two eigenvalues.    

\subsection{Dephasing Environment}

We now look at the entanglement evolution of the three qubit $X$-states with no
interaction between the qubits, in an independent qubit dephasing environment 
noting the exhibition of ESD with respect to the negativity. The independent qubit dephasing 
environment is fully described by the Kraus operators
\begin{equation}
K_1 = \left(
\begin{array}{cc}
1 & 0 \\
0 & \sqrt{1-p} \\
\end{array}
\right); \;\;\;\;
K_2 = \left(
\begin{array}{cc}
0 & 0 \\
0 & \sqrt{p} \\
\end{array}
\right),
\end{equation} 
where the dephasing parameter $p$ can also be written in a time-dependent fashion, 
$p = 1-\exp(-\kappa t)$. When all three qubits undergo dephasing we have eight 
Kraus operators each of the form 
$A_l = (K_i\otimes K_j\otimes K_k)$ where $l = 1,2,...,8$ and $i,j,k = 1,2$.

\begin{figure}[t]
\includegraphics[width=5cm]{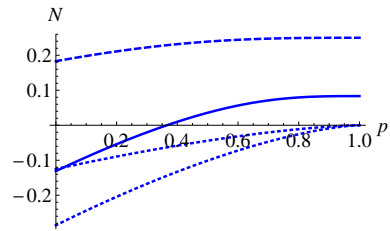}
\caption{(Color online) Eigenvalues of partially transposed $X$-state density matrix with
non-zero elements $a_1 = \frac{5}{12}, a_2 = \frac{3}{12}, b_1 = \frac{1}{4}, 
b_2 = \frac{1}{12}, c_1 = \frac{2}{7}$, and $c_2 = \frac{1}{8}$. When taking 
the partial transpose with respect to the third qubit there are
two `negative' eigenvalues which can be calculated from Eq. \ref{ZEigs}: 
$\frac{1}{6}-\frac{1}{84}\sqrt{625-576p(3+p(p-3)}$ and $\frac{1}{24}(8-\sqrt{13-9p(3+p(p-3))})$.
Using Eq. \ref{pEq1} we can see that only the first of these eigenvalues can be negative and
this  for $p < .366$. Thus, the state exhibits ESD with respect to $N_3$. These two eigenvalues are shown in the 
plot (solid and dashed line respectively). In addition, the two negative eigenvalues that constitute 
$N_1$ and $N_2$ are shown (dotted lines) neither of these go to zero in finite time. Thus, the state
exhibits ESD with respect to the tri-partite negativity at the same value as exhibited for $N_3$. }
\label{fig1}
\end{figure}

The effect of a dephasing environment of strength $p$ on an $X$-state is to 
reduce the anti-diagonal elements of the density matrix by a factor
$(1-p)^{3/2}$ while leaving the diagonal elements constant. To calculate the 
negativity we look at the eigenvalues of the $X$-state after taking 
the partial transpose with respect to the desired subsystem. The relevant 
eigenvalues are now of the form: 
\begin{equation}
\frac{1}{2}\left(a_j+b_j-\sqrt{(a_j-b_j)^2+4|c_i|^2(1-p)^3}\right).
\label{ZEigs}
\end{equation}
The eigenvalues go to zero when:
\begin{equation}
p = 1-\frac{(a_j b_j)^{1/3}}{|c_i|^{2/3}}.
\label{pEq1}
\end{equation} 
Based on the above, it is easy to see that ESD with respect to negativity is not 
exhibited by $X$-states made up of single GHZ-type states (with non-zero elements 
$a_j + b_j =1$, and $c_j$): the negativity with respect to any one qubit, and thus the 
tri-partite negativity as well, is simply $-|c_j|(1-p)^{3/2}$. When the $X$-state is 
a mixture of GHZ type states ESD may be exhibited.
Fig. \ref{fig1} shows a sample $X$-state that is the sum of two GHZ-type 
states that exhibits ESD with respect to the negativity of the third qubit, $N_3$. 
However, the state does not exhibit ESD with respect to $N_1$ and $N_2$, they are negative for any 
value of $p$. The state thus exhibits ESD with respect to the tri-partite negativity at the same 
dephasing strength as $N_3$. For stronger dephasing no tri-partite entanglement is detected. 

The above can be compared to the experimental detection capabilities of entanglement witnesses.
Appropriate entanglement witnesses for $X$-states are of the sort:
\begin{equation}
W_k = \frac{3}{4}\openone - |GHZ(k)\rangle\langle GHZ(k)|
\end{equation}
where $|GHZ(k)\rangle = \frac{1}{\sqrt{2}}(|k\rangle+|\overline{k}\rangle)$. 
For an $X$-state consisting of a single GHZ-type
state the entanglement witness $W_k$ gives $Tr[W_k\rho] = \frac{1}{4}(1-4(1-p)^{3/2}|c_k|)$.
Thus, $W_k$ loses its ability to detect entanglement at dephasing strength 
$p = 1-\frac{1}{2^{4/3}|c_k|^{2/3}}$. 
The maximum occurs for $|c_k| = 1/2$ in which case the entanglement is no longer detected 
at $p = 1-\frac{1}{2^{1/3}}$. For general $X$-states the entanglement witnesses give the 
following:
\begin{eqnarray}
\rm{Tr}[W_{\ell}\rho] &=& \frac{1}{4}(3(a_i+b_i+a_j+b_j+a_k+b_k) \nonumber\\
			    &+& a_{\ell}+b_{\ell}-4(1-p)^{3/2}|c_\ell|).
\end{eqnarray}
This can be solved for the exact value of $p$ at which the entanglement is no longer detected:
\begin{equation}
p = 1-\frac{(3(a_i+b_i+a_j+b_j+a_k+b_k)+a_{\ell}+b_{\ell})^{2/3}}{2^{4/3}|c_{\ell}|^{2/3}}.
\end{equation}
I note that which of the above witnesses is most sensitive may depend on the intial state and 
the decoherence strength and can be determined via a minimization process. What is important 
is that the witnesses detect purely tri-partite entanglement that does not include biseparable 
but not completely separable entanglement. 

\subsection{Depolarizing Environment}

I now look at an independent qubit depolarizing environment and, as above, explore the 
entanglement evolution of the three qubit $X$-states. The Kraus
operators for this environment are:
\begin{equation}
K_1= \sqrt{1-\frac{3p}{4}}\openone,
K_w = \frac{\sqrt{p}}{2}\sigma_w, 
\end{equation}
where $\sigma_w$ are the Pauli spin operators,
$w = x,y,z$ and $p$ is now the depolarizing strength. The depolarizing environment 
affects both the anti-diagonal and diagonal elements of the density matrix. The anti-diagonal 
elements are simply reduced by a factor of $(1-p)^3$. The diagonal element $a_i$ becomes:
\begin{eqnarray}
a_i^{\prime} &=& a_i(1-\frac{3p}{2}+\frac{3p^2}{4}-\frac{p^3}{8}) \nonumber\\
		 &+& (a_j+a_k+b_{\ell})(\frac{p}{2}-\frac{p^2}{2}+\frac{p^3}{8}) \nonumber\\
		 &+& b_i\frac{p^3}{8}+(a_{\ell}+b_j+b_k)(\frac{p^2}{4}-\frac{p^3}{8})
\end{eqnarray}
where if $(i,\ell) = (1,4)$, $(j,k) = (2,3)$ and vice versa. For the $b_i^{\prime}$ elements
simply replace each term $b_l$ with $a_l$ and each $a_l$ with $b_l$, for $l = 1,...,4$. 
Since the decohering environment preserves the $X$ shape of the density matrix the eigenvalues
of the partially transposed density matrix follow Eq. \ref{XEigs} and the critical value of $p$
for which a given eigenvalue goes from negative to positive can be analytically determined. 

\begin{figure}[t]
\includegraphics[width=5cm]{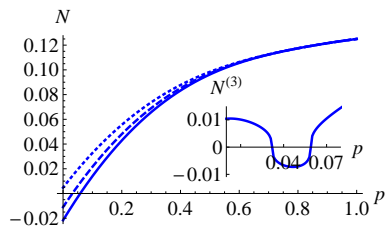}
\caption{(Color online) Eigenvalues of partially transposed $X$-state density matrix with
non-zero elements $a_1 = \frac{1}{8}, a_2 = \frac{1}{8}, a_3 = \frac{1}{8}, a_4 = \frac{1}{16}, 
b_1 = \frac{1}{8}, b_2 = \frac{1}{8}, b_3 = \frac{3}{16}, b_4 = \frac{1}{8}, c_1 = \frac{1}{12}, 
c_2 = \frac{1}{9}, c_3 = \frac{1}{10}$, and $c_4 = \frac{2}{25}$. The most negative eigenvalue
is shown for the partial transpose of the initial $X$-state density matrix taken with 
respect to the first (dotted line), second (solid line), and third (dashed line) qubit.
The state exhibits ESD with respect to the latter two negativities at dephasing values 
$p \simeq .0585$ and $.0317$, respectively. The inset shows the tri-partite negativity. 
Note that this tri-partite entanglement measure goes from positive (indicating the lack of 
$N^{(3)}$ entanglement) to negative (indicating the presence of GHZ distillable entanglement), 
back to positive. This entanglement behavior arises from the fact that the single qubit 
negativities cross zero at different decoherence strengths. 
}
\label{fig2}
\end{figure}

When the initial density matrix is composed of only one GHZ-type state eigenvalues of the 
partially transposed density matrix are the same for each qubit and the state exhibits ESD 
at the same depolarizing strength for all $N_j$ and $N^{(3)}$.
When the $X$-state density matrix is a mixture of multiple GHZ-type states ESD
may be exhibited with respect to specific negativity measures at different 
depolarizing strengths. Figure \ref{fig2} 
shows the lowest eigenvalue of the partially transposed density matrix with respect to each 
of the three qubits for a sample $X$-state composed of a mixture of GHZ-type states 
as a function of decoherence strength. One eigenvalue is always positive 
({\it i.e.} indicating zero negativity) and two of the eigenvalues cross zero 
({\it i.e.} the state undergoes ESD with respect to the single qubit negativities) 
at different decoherence strengths. For low values of $p$ two of the lowest eigenvalues are negative 
demonstrating the presence of entanglement. However, there is no measurable GHZ distillable tri-partite 
entanglement as measured by $N^{(3)}$. For slightly higher values of $p$ there is a small region 
for which only one of the eigenvalues is negative. Now $N^{(3)}$ becomes negative showing 
a sudden {\it birth} of (GHZ distillable) tri-partite negativity. As $p$ increases further, the state exhibits
ESD with respect to all single qubit negativities and $N^{(3)}$ (and $N_2$) becomes positive. 
This sort of $N^{(3)}$ behavior indicates that there is only a small region of decoherence strengths 
(which does not include $p = 0$) for which we can be sure there exists GHZ-distillable entanglement.
Such behavior, going from positive to negative and back, cannot occur when the $X$-state is composed 
of only two GHZ-type states. This is because two of the single qubit negativites are equal, the partial 
trace with respect to two of the qubits give the same set of eigenvalues. The sign of $N^{(3)}$ is thus 
determined solely by the eigenvalues of the partially transposed state with respect to the third qubit. An example of 
a state exhibiting the sudden birth of $N^{(3)}$ entanglement followed by an exhibition of ESD 
is portrayed in the inset of Fig. \ref{fig2}.

To test for the presence of purely tri-partite entanglement in the depolarizing system we look 
at the expectation value of the state with an entanglement witness. Using the witnesses 
defined above we note that witness $W_j$ for a state depolarized with a strength $p$ gives:
\begin{eqnarray}
\rm{Tr}[W_j\rho] &=& \frac{1}{8}(p^2-2p+6)(\sum_{i \neq j} a_i+b_i) \nonumber\\
		     &-& \frac{1}{8}(3p^2-6p-2)(a_j+b_j)+(p-1)^3c_j.
\end{eqnarray}
This equation can then be solved for the critical decohernce strength at which entanglement 
will no longer be detected.

\section{$X$-States with More Qubits}

The particular matrix structure of the $X$-state allows us to extend our results beyond three 
qubits. A matrix with non-zero elements in an $X$ shape can be block diagonalized with blocks of 
size $2\times 2$. Assuming the $X$-matrix elements along the diagonal are $d_1,...,d_N$,
and the elements along the anti-diagonal are $e_1,...,e_N$ (starting at the top right), the $m$th 
$2\times 2$ block along the diagonal has elements:
\begin{equation}
A_m = \left(
\begin{array}{cc}
d_m & e_m \\
e_{N-m+1} & d_{N-m+1} \\
\end{array}
\right); 
\label{BDiag}
\end{equation}
where $1 \leq m \leq N/2$. Thus, the eigenvalues of any dimension $X$-shaped matrices 
are simply the eigenvalues of the $2\times 2$ blocks which are 
\begin{equation}
\frac{1}{2}(d_m+d_{N-m+1}\pm\sqrt{(d_m-d_{N-m+1})^2+4e_me_{N-m+1}}).
\label{Eig2}
\end{equation} 

When calculating the negativity a partial transpose of the density matrix must be taken. This has the effect
of rearranging only the elements along the anti-diagonal while preserving the $X$ shape. Thus, the eigenvalues
of the partial transpose of an $X$-state of any dimension have the form of Eq. \ref{XEigs} and the negativity 
is easily calculated.

\section{Three-Qubit Concurrence}
In this section I derive an explicit expression for a lower bound of 
three qubit mixed state concurrence as defined in \cite{LFW} for $X$-states. 
A general expression for a lower bound on the three-qubit concurrence is:
\begin{equation}
\tau_3 = \sqrt{\frac{1}{3}\sum^6_{\ell=1}\left[(C_{\ell}^{12|3})^2+(C_{\ell}^{13|2})^2+(C_{\ell}^{23|1})^2\right]}.
\end{equation}
Each of the three bi-partite concurrence terms $C^{ij|k}_\ell$ is given as the sum of the six terms:
\begin{equation}
C_{\ell} = \rm{max}\{0,\sqrt{\lambda^1_\ell}-\sqrt{\lambda^2_\ell}-\sqrt{\lambda^3_\ell}-\sqrt{\lambda^4_\ell}\},
\end{equation}
where $\lambda^l_\ell$ are the non-zero eigenvalues of $\tilde{\rho} = \rho S_\ell^{ij|k}\rho^*S_\ell^{ij|k}$ 
in descending order. The operators $S_\ell^{ij|k}$ are given by $S_\ell^{ij|k} = L_\ell^{ij}\otimes L_0^k$ 
where $L_\ell^{ij}$ is one of six generators of the group SO(4) operating on qubits $i,j$, and $L_0^k$ 
is the generator of SO(2), the Pauli matrix $\sigma_y$, operating on qubit $k$. This lower bound on mixed 
state concurrence has been calculated for some simple $X$-states in \cite{SF}. Here I look to extend these 
results and note where the lower bound goes to zero. Once the lower bound does go to zero, there is no longer
a guarantee that entanglement is present.

For initial density matrices that are $X$-states only six of the 18 contributing terms to the three 
qubit concurrence are non-zero. More specifically, only the two SO(4) generators with elements on 
the anti-diagonal contribute to each of the three bi-partite concurrences. I will refer to the 
SO(4) generator with anti-diagonal $(-1,0,0,1)$ as $L_1^{ij}$, and the SO(4) generator with 
anti-diagonal $(0,-1,1,0)$ as $L_2^{ij}$ . For $X$-states the four eigenvalues that make up 
each of the six terms are of the form:
\begin{equation}
\lambda^{ij|k}_{\ell,m_{\pm}} = a_mb_m+|c_m|^2\pm2\sqrt{a_mb_m|c_m|^2}
\end{equation}
for two different values of $m$. For the partition $12|3$ and SO(4) generator $\ell = 1$ the contributing
terms have $m = 1,2$. For the generator $\ell = 2, m = 3,4$. Similarly, for the partition $23|1$ we find
$\ell = 1, m = 2,3$ and $\ell = 2, m = 1,4$. Finally, for the $13|2$ partition $\ell = 1, m = 2,4$ and 
$\ell = 2, m = 1,3$. Given these eigenvalues the three-concurrence can be easily computed. 

\subsection{Dephasing Environment}

For $X$-states composed of only one GHZ-type state an exact calculation for $\tau_3$
in a dephasing environment yields the maximum between 0 and:
\begin{equation}
\sqrt{a_i-a_i^2+c_i(-\omega_i+2\gamma_i^{\frac{1}{2}})} - \sqrt{a_i-a_i^2+c_i(-\omega_i-2\gamma_i^{\frac{1}{2}})}
\label{DephaseEig}
\end{equation}
where,
\begin{eqnarray}
\omega_i &=& c_i(p-1)^3 \nonumber\\
\gamma_i &=& a_i(a_i-1)(p-1)^3.
\end{eqnarray}
Eq. \ref{DephaseEig} goes to zero only in the limit of $p\rightarrow 1$. Thus, the lower bound cannot
go to zero for a GHZ-type state, some entanglement will always be present. In fact, the lower bound
cannot go to zero unless the $X$-state is composed of four GHZ-type states. This is because $\tau_3$ 
is a summation of terms and can go to zero only if each term goes to zero. Each one of these 
(six) terms consists of four eigenvalues, two for each of two $m$ values. If the eigenvalues
of one of the $m$ values are zero (which will happen if $c_i$ and $a_i$ or $b_i = 0$) the term 
will have the form of Eq. \ref{DephaseEig}. Thus, that term, if not initially zero, will remain 
non-zero until $p = 1$. This behavior is demonstrated in Fig. \ref{fig3} and should be contrasted 
with the negativity and tri-partite negativity measures. The negativity of $X$-states can go to 
zero in a dephasing channel when the $X$-state is composed of only two GHZ-type 
states. The reason for this is that the negativity can be defined with respect to only one of the
qubits (for example $N_3$) which may go to zero while negativity measures with respect to the other
qubits do not. The three qubit concurrence, however, is a sum over all terms and therefore cannot go
to zero unless each bi-partite concurrence term goes to zero.

\subsection{Depolarizing Environment}

\begin{figure}[t]
\includegraphics[width=3.5cm]{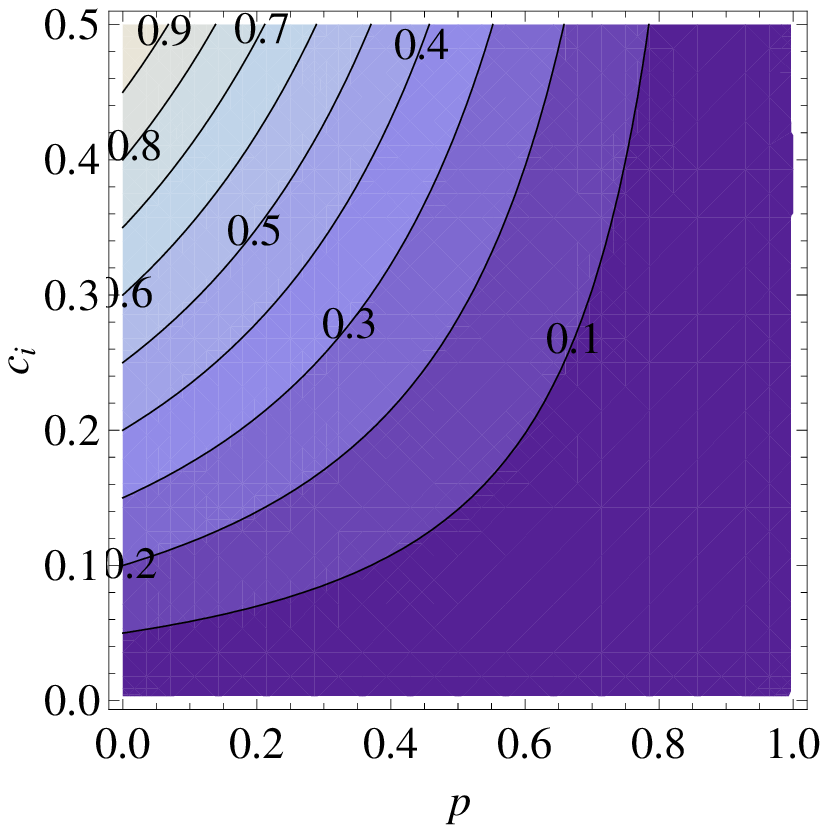}
\includegraphics[width=3.5cm]{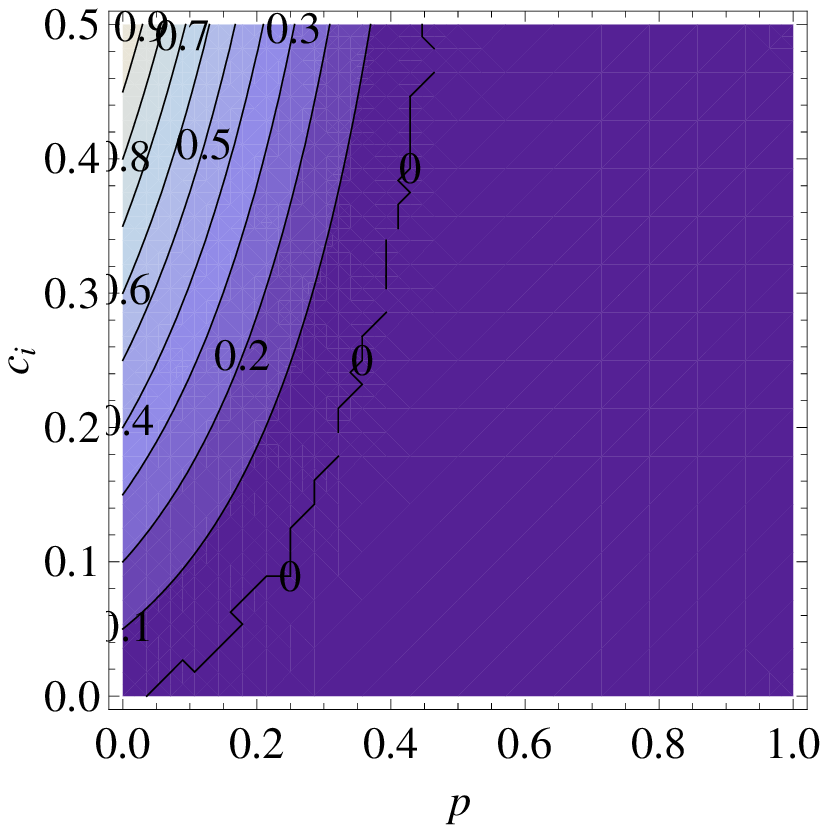}
\includegraphics[width=3.5cm]{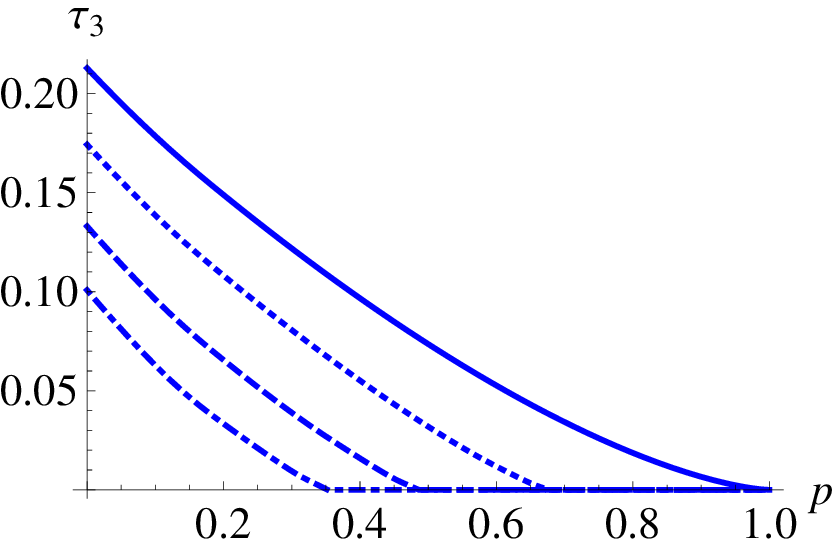}
\caption{(Color online) Top row: Three qubit concurrence for $X$-states composed of one GHZ-type
state with $a_i = b_i = 1/2$. Left: in a dephasing environment $\tau_3$ never goes to zero. 
Right: in a depolarizing environment $\tau_3$ goes to zero when $\frac{1}{4}(p(p-2)+8c_i(1-p)^3)=0$.
Bottom: example of three-qubit concurrence evolution as a function of dephasing strength.
The chosen $X$-state density matrix is: $a_1 = \frac{3}{8}-\epsilon, a_2 = \frac{1}{4}, a_3 = \epsilon, 
a_4 = \frac{1}{16}, b_1 = \frac{1}{16}, b_2 = \frac{1}{16}, b_3 = \frac{1}{16}, b_4 = \frac{1}{8}, 
c_1 = \frac{3}{25}, c_2 = \frac{1}{9}, c_3 = \frac{\epsilon}{2}$, and $c_4 = \frac{1}{12}$. $\tau_3$ is 
shown for $\epsilon = 0$ (solid line), $\epsilon = 1/128$ (dotted line), $\epsilon = 1/32$ (dashed
line), and $\epsilon = 1/16$ (chained line). Note that when $a_3 = 0$ $\tau_3$ does not go to zero.
Adding any finite amount to $a_3$ allows $\tau_3$ to go to zero. }
\label{fig3}
\end{figure}

The bound on the three-qubit concurrence for $X$-states composed of one GHZ-type state in a 
depolarizing environment gives the maximum between 0 and:
\begin{equation}
\frac{q_i}{4}-\sqrt{\omega^2_i-\frac{\gamma^{\prime}_i}{64}-
		\frac{\sqrt{\omega^2_i\gamma^{\prime}_i}}{4}} + \sqrt{\omega^2_i-\frac{\gamma^{\prime}_i}{64}+
		\frac{\sqrt{\omega^2_i\gamma^{\prime}_i}}{4}}
\end{equation}
where,
\begin{eqnarray}
q_i &=& (p-2)p\sqrt{4(p-1)^2(a_i-a_i^2)-p(p-2)} \\
\gamma^{\prime}_i &=& (a_i(p-2)^3+(a_i-1)p^3)((a_i-1)(p-2)^3+a_1p^3). \nonumber
\end{eqnarray}
$\tau_3$ for this state is shown in Fig. \ref{fig3} for an initial state $a_i = b_i = 1/2$. 
For the depolarizing environment $\tau_3$ can go to zero even for $X$-states composed of 
only one GHZ-type state. 

The matrix $\tilde{\rho}$ for a $X$-state density matrix with any of the six SO(4) 
generators retains its $X$ shape, having four diagonal and four anti-diagonal elements
(and thus have only four non-zero eigenvalues as noted in \cite{LFW}). The $X$ is 
`balanced' when $S_\ell^{ij|k}$ is anti-diagonal. I use the term `balanced' as follows: 
any diagonal $X$-matrix element $d_m$ that is non-zero has a non-zero counterpart 
element $e_m$, where I have used the notation of Eq. \ref{BDiag}.
An unbalanced $X$-matrix will have non-zero elements whose counterparts are zero. 
When a balanced $X$-matrix is block diagonalized non-zero $2\times2$ blocks will have 
four non-zero elements leading to non-degenerate eigenvalues like those of Eq. \ref{Eig2}. 
Block diagonalized unbalanced $X$-matrices will have a zero in one of the off-diagonal elements 
of the $2\times2$ diagonal block as can be noted from Eq. \ref{BDiag}. In the 
unbalanced case the eigenvalues are then simply the diagonal elements of the block
which are of the form $a_ib_j,a_ib_j,a_kb_l,a_kb_l$. These elements (eigenvalues) are 
degenerate and thus these bi-partite concurrence terms equal zero.  

\subsection{$n$-Qubit Concurrence}

As mentioned above, only the two SO(4) generators with elements on the anti-diagonal, 
contribute to the three qubit concurrence. This is because only these two 
generators have anti-diagonal elements which lead to balanced $X$-matrices 
$\tilde{\rho}$. The other generators lead to unbalanced $\tilde{\rho}$ matrices whose
eigenvalues are simply its diagonal elements. The eigenvalues are each doubly degenerate 
meaning that these terms will not contribute to the concurrence. 
The above allows us to simplify calculations for higher qubit concurrences of 
$X$-states. The only terms necessary to calculate are those that utilize anti-diagonal
$S_\ell^{ij...|k\ell...}$ matrices. Thus, for four qubits there would be four terms from the 
$\rm{SO(4)}\otimes\rm{SO(4)}$ generators for each of the three balanced partitions 
(two qubits on each side of the partition) and an additional four terms from the 
$\rm{SO(8)}\otimes\rm{SO(2)}$ generators for each of the four unbalanced partitions 
(partitions of three and one qubit). This gives a total of 28 terms which 
should significantly simplify these calculations. 

\section{Conclusions}

In this paper I have studied the entanglement dynamics for three qubit $X$-states in 
both dephasing and depolarizing environments. To do this I have analytically calculated the eigenvalues 
of partial transposes of the $X$-states which allows for easy determination of the negativity. Since the 
dephasing and depolarizing environments retain the density matrix $X$ shape one can calculate which initial 
states will exhibit ESD with respect to the negativity measures and at what decoherence strength. 
I noted that the tri-partite negativity, a tri-partite entanglement measure which is sufficient to 
ensure GHZ distillability, can exhibit non-standard behavior for certain $X$-states: its appearance 
only at non-zero decoherence strength followed by its sudden disappearance. In addition, I explored 
the detection capability of entanglement witnesses sensitive to tri-partite entanglement. As with 
the negativity, the expectation value of the $X$-state with respect to the entanglement witness can 
be solved analytically and are vital in assessing potential experimental studies.
These results are extended to systems made of arbitrary numbers of qubits. Finally, I analytically 
solved for the relevant terms of a lower bound on the three-qubit concurrence for an $X$-state, demonstrated 
when it goes to zero in dephasing and depolarizing environments. This method may be useful for calculating 
concurrences for larger numbers of qubits.


It is a pleasure to thank G. Gilbert and S. Pappas for helpful feedback 
and acknowledge support from the MITRE Innovation Program under MIP grant \#20MSR053.

\appendix
\section{General X-States}
As mentioned in the main part of the paper, a previously studied set of states with an $X$ 
shaped density matrix is the generalized GHZ-diagonal states \cite{A}. In this Appendix 
I prove that generalized GHZ-diagonal states do not include all possible X-states by constructing an explicit 
state with an X-shaped density matrix that is not part of the aforementioned set. 

Generalized GHZ-diagonal states with $n$ qubits have the form
\begin{equation}
\rho = \sum_{k = 0}^{N-1}\lambda_k^+|\psi_k^+(\alpha,\beta)\rangle\langle\psi_k^+(\alpha,\beta)|+\lambda_k^-|\psi_k^-(\alpha,\beta)\rangle\langle\psi_k^-(\alpha,\beta)|,
\end{equation}
where $N = 2^n$ is the Hilbert space dimension. The density matrix elements of these states using 
the notation of Eq.~\ref{xstate} are as follows: 
\begin{eqnarray}
a_j &=& |\alpha|^2(\lambda_k^{+}+\lambda_k^{-}) + |\beta|^2(\lambda_{\overline{k}}^{+}+\lambda_{\overline{k}}^{-}) \\
b_j &=& |\beta|^2(\lambda_k^{+}+\lambda_k^{-}) + |\alpha|^2(\lambda_{\overline{k}}^{+}+\lambda_{\overline{k}}^{-}) \\
c_j &=& \alpha\beta^*(\lambda_k^{+}-\lambda_k^{-}) + \alpha^*\beta(\lambda_{\overline{k}}^{+}-\lambda_{\overline{k}}^{-}) 
\end{eqnarray}
for $1 < j < N/2$ and $k = j - 1$.

We now construct an $X$-state that is not a generalized GHZ-diagonal state. 
Let us set $c_j = 0$. There are then three possible solutions for Eq.~A4:

\begin{description}
\item [$\alpha = 0$ or $\beta = 0$] {~\\
If either of these is true then all other $c_m$ for $m \neq j$ must also equal zero. }

\item [$\lambda_{k}^+ = \lambda_{k}^-$ \emph{and} $\lambda_{\overline{k}}^+ = \lambda_{\overline{k}}^-$] {~ \\
If this is true $a_j = 2(|\alpha|^2\lambda_{k} + |\beta|^2\lambda_{\overline{k}})$ and 
$b_j = 2(|\beta|^2\lambda_{k} + |\alpha|^2\lambda_{\overline{k}})$ where 
$\lambda_{k} = \lambda_{k}^+ = \lambda_{k}^-$. Therefore, if in addition 
$b_j = 0$ (which would require $\lambda_{k} = \lambda_{\overline{k}} = 0$), 
$a_j$ must equal zero.} 

\item [$(\lambda_{k}^{+}-\lambda_{k}^{-}) = \frac{\alpha^*\beta}{\alpha\beta^*}(\lambda_{\overline{k}}^{-}-\lambda_{\overline{k}}^{+})$] {~\\
Let $\alpha^*\beta = re^{i\theta}$ where $r,\theta$ are real. Then the fraction $\frac{\alpha^*\beta}{\alpha\beta^*} =
e^{2i\theta}$. As mentioned in the main part of the paper, the coefficients $\lambda$ are all real forcing
$e^{2i\theta}$ to be real and $\theta = 0,m\pi$ for all integers $m$. Therefore, $\alpha^*\beta$ and 
$\alpha\beta^*$ must both be purely real or purely imaginary.}

\end{description}

We can now explicitly construct a two qubit $X$-state that is not part of the set of generalized 
GHZ-diagonal states by setting $c_1 = 0$ and violating each of the three conditions listed above. 
Such a state can have the form:
\begin{equation}
\rho_C = \left(
\begin{array}{cccc}
a_1 & 0 & 0 & 0 \\
0 & a_2 & re^{i\phi} & 0 \\
0 & re^{-i\phi} & b_2 & 0 \\
0 & 0 & 0 & 0 \\
\end{array}
\right).
\label{vstate}
\end{equation}
where $r,\theta$ are real and $r^2 < a_2b_2$ guarantees the density matrix has 
positive eigenvalues. In addition, $\rho_C$ must be trace 1, $a_1+a_2+b_2 = 1$, 
and its purity must be $a_1^2 + a_2^2 + b_2^2 +2r^2 \leq 1$. 

In Eq. \ref{vstate} $c_1 = 0$, yet $c_2 \neq 0$, indicating that $\alpha,\beta\neq 0$.
Furthermore, $b_1 = 0$ while $a_1$ does not, indicating that $\lambda_0^+ = \lambda_0^-$
and $\lambda_3^+ = \lambda_3^-$ cannot both be true. Finally, $c_2 = re^{i\phi}$ need not be 
real nor purely imaginary indicating that 
$(\lambda_0^{+}-\lambda_0^{-}) \neq \frac{\alpha^*\beta}{\alpha\beta^*}(\lambda_3^{-}-\lambda_3^{+})$.
Thus, the state $\rho_C$ is not part of the set of generalized GHZ-diagonal states though
it certainly is an $X$-state.


\end{document}